# Photo-Thermal Neural Excitation by Extrinsic and Intrinsic Absorbers: A Temperature-Rate Model

Nairouz Farah[1], Inbar Brosh[1], Christopher R. Butson[2] and Shy Shoham[1*]

[1]Faculty of Biomedical Engineering,
Technion - Israel Institute of Technology
Haifa 32000, ISRAEL
[2]Departments of Neurology & Neurosurgery
Medical College of Wisconsin
Milwaukee, WI, USA

* Corresponding Author
Shy Shoham, Ph.D.
Faculty of Biomedical Engineering,
Technion - Israel Institute of Technology,
Haifa, Israel Technion City, Haifa 32000
Israel
Phone: +972-4-829-4125
Fax: +972-4-829-4599
Email: sshoham@bm.technion.ac.il

**Infrared neural stimulation (*INS*) pulses at water-absorbed mid-IR wavelengths could provide a non-invasive and safe modality for stimulating peripheral and cranial nerves and central nervous system neurons. The excitation mechanism underlying INS activation is thought to be mediated by photo-thermal tissue transients, which can also potentially be induced using extrinsic absorbers (Photo-Absorber Induced Neural-Thermal Stimulation or *PAINTS*). The specific biophysical effect of photo-thermal transients on target neurons has yet to be determined and quantitatively characterized.**

**Here, we propose and study a model for thermally-induced neural stimulation where temperature changes induce a depolarizing transmembrane current proportional to the temperature rate of change. Our model includes physical calculations of the temperature transients induced by laser absorption and a biophysical model of the target cells. Our results indicate that stimulation thresholds predicted by the model are in good agreement with empirical data obtained in cortical cell cultures using extrinsic micro-particle absorbers (*PAINTS*) as well as with earlier results on auditory neuron stimulation using INS. These results suggest a general empirical-law for photo-thermal interactions with neural systems, and could help direct future basic and applied studies of these phenomena.**

# 1. Introduction

Recent years have seen the emergence of *applied* optical neurostimulation, driven by the introduction of two promising new optical technologies for activating nerve cells: light-sensitive ion channels and pumps[1] ("optogenetic stimulation") and direct INS stimulation[2, 3]. Despite the tremendous promise of optogenetic photo-stimulation tools, stable transduction is currently only obtainable through viral transfection, and may face substantial barriers currently faced by other gene-therapy approaches, prior to its clinical application. An alternative, potentially more direct and simple path to light-mediated neuro-stimulation in various neural tissues may become possible using mid IR laser pulses tuned to water-absorption peaks around 2μm. Effective stimulation using this approach was demonstrated, for example, in peripheral nerve[2], cranial nerves (facial[4], auditory[5] & vestibular[6]), central auditory system *in vivo*[7] and *in vitro* CNS neuronal preparation[8]. Similarly, we recently reported neural stimulation based on photo-thermal excitation of exogenous photo-absorbers, an approach tentatively termed Photo Absorber Induced Neural-Thermal Stimulation or *PAINTS*[9-11].

The experimental work of Wells *et al.*[12] studied three probable mechanisms for INS excitation (electric-field mediated, photo-mechanical and photo-thermal), concluding that the most likely mechanism is photo-thermal, resulting from the absorption of IR illumination by the tissue's water content. Indeed, there are a number of well known temperature-dependent biophysical effects that could potentially be involved in mediating photo-thermal effects of a heating pulse: a decrease in the Nernst potential, an increase in ion channel conductance, an increase of open state transitions of the sodium channels and a thermal volumetric expansion. Another hypothesis is that INS stimulates the naturally temperature sensitive channels TRPV[3]. However, intracellular recordings from afferent primary neurons during INS[13], demonstrated evoked depolarizations with a near-uniform distribution of responsive neurons, apparently rejecting this possibility. Recent work by Teudt *et al.*[14] suggests that thermal volumetric expansion may play a more significant role in auditory neuron stimulation than previously thought[3]. Finally, an overview of biophysical photo-stimulation mechanisms by Sjulson and Miesenbock[15] notes briefly that the likely mechanism for thermal stimulation is probably depolarization mediated by temperature-dependent gating of voltage gated sodium channels. Sjulson and Miesenbock suggest that this thermal effect can be augmented by expressing TRPV channels in the targeted neurons.

Clearly, the development of a specific, quantitative and predictive understanding of thermo-bioelectrical coupling mechanisms in photo-thermal neuro-stimulation is not only significant scientifically, but could also greatly assist the design of future stimulation devices, configurations and applications, as it has for example in the case of electrical stimulation[16].

To address this open challenge we investigated predictions from a 'temperature-rate' hypothesis wherein temperature transients induce an inward transmembrane stimulation current which is proportional to their time derivative ($I \propto dT/dt$). A temperature-rate model can intuitively explain why thermally mediated excitation effects weren't previously observed under slow thermal changes while short pulses appear very effective in driving neural excitation, and why excitation appears to be independent of baseline temperatures[12]. Moreover, an inverse temperature-rate model (where $I \propto -dT/dt$) was derived by Barnes[17] from Nernst equilibrium relations, in his attempt to explain the thermal effects of electromagnetic radiation deposition on neurons. The new empirical model is tested in a general setting, by applying it to empirical data from experiments involving localized photo-thermal stimulation using both extrinsic and intrinsic photo-absorbers. The first test dataset is obtained from new *PAINTS* experimental results where we measured neuronal responses to thermal transients near microscopic photo-absorbers distributed in the vicinity of the cultured cortical neurons and illuminated by laser pulses. In these experiments the induced thermal transients have a highly localized nature and fast temporal dynamics, resulting in activation patterns with high spatiotemporal resolution. The second test dataset was based on the detailed experimental measurements by Izzo and colleagues of INS cochlear stimulations[18, 19]. To complete the quantitative description of an entire stimulation experiment, we compute the temperature dynamics using analytical solutions for the photo-thermal heat equation, and use a detailed biophysical model to calculate the neuronal response thresholds to the induced membrane currents. Our results indicate very good agreement between the predictions of the temperature-rate model and a range of experimental results.

# 2. Methods

## 2.1 Cell Culture PAINTS Experiments

*Cell culture preparation and calcium imaging*: Cortical tissue was obtained from 0-2 days old Sprague-Dawley rats. Cells were plated on PEI (Polyethyleneimine) coated coverslips, onto which micron-scale absorbers were first dispersed. The preparation was stained with the calcium indicator OGB-1 (Invitrogen – Molecular Probes). To each 50µg vial of the indicator, 8µl of DMSO (Sigma-Aldrich), 2µl 20% Pluoronic in DMSO (F-127, Biotium Inc), and 90µl of cell medium were added. Dye solution was then added to the cell culture to a final concentration of 2.64µM (20µl/3ml) for an incubation period of 35-40 minutes after which dye wash out was performed. Calcium transients in the stained culture were imaged through an inverted microscope (TE2000U, Nikon) using a CCD camera (C8484-05G, Hamamatsu Photonics), at a frame rate of 10 frames/sec.

*PAINTS Experiments*: The photo-thermal stimulation experiments were performed 1-3 weeks after plating of the primary culture. To induce thermal activation of the cortical neurons, high intensity light patterns were directed onto particles in the vicinity of the cells using a computer based holographic system developed by our group[20]. Stimulation power thresholds for varying pulse durations were measured.

## 2.2 Computational Modelling

Detailed computational models that capture both physical and biophysical properties of the experimental setting were developed to describe laser-induced temperature transients and their effect on nerve cells. In these models we tested a temperature-rate hypothesis wherein a temperature change leads to an effective transmembrane current proportional to the derivative of the temperature w.r.t. time.

*PAINTS*

To calculate the tissue temperatures generated by PAINTS experimental conditions, we used an analytical solution to the heat equation around a discrete absorber[21]:

$$
\begin{cases}
T(r,t) = \dfrac{P}{4\pi kr} erfc\left(\dfrac{r}{\sqrt{4\alpha t}}\right) & t < \tau \\
T(r,t) = \dfrac{P}{4\pi kr}\left[erfc\left(\dfrac{r}{\sqrt{4\alpha t}}\right) - erfc\left(\dfrac{r}{\sqrt{4\alpha\left(t-\tau\right)}}\right)\right] & t \geq \tau
\end{cases} \quad (1)
$$

Where $P$[W] is the pulse's power, $\tau$[sec] is the pulse duration, $\alpha=1.4*10^{-7}$ [$m^2$/sec] is the thermal diffusivity, $k$=0.6 [W/mK] is the thermal conductivity and $r$[m] is the distance from the centre of the absorber.

To simulate the biophysical properties of a typical cortical cell in our culture (including the fundamental morphological and electrical properties) we used a NEURON biophysical model of a layer 5 pyramidal neuron, developed by Fleidervish *et al.*[22].

*Auditory Nerve IR stimulation*

We also simulated the experimental conditions reported by Izzo, Richter and colleagues in their studies investigating the dependence of optical auditory neuron stimulation on laser wavelength and pulse duration[18, 19]. The physical parameters and pulse durations are summarized in Table 1. Physical modelling involved calculating the resulting temperature transients for the given laser beam geometry and the tissue's thermal and physical properties. The temperature transient was calculated analytically following the derivation detailed by Welch and Van Gemert[21] for a tissue illuminated by a Gaussian shaped radial beam. Assuming that the tissue at hand is a homogenously absorbing infinite layer having an initial temperature rise of zero, the temperature change can be readily calculated by (numerically) solving the following integral:

$$
\begin{aligned}
T\left(r,z,t\right) &= \frac{\mu_a E_0}{2\rho c}\exp\left(-\mu_a z\right)\times \\
&\int_0^t u\left(\tau+t'-t\right)\frac{\exp\left(\mu_a{}^2\alpha t' - \dfrac{2r^2}{w_l{}^2+8\alpha t'}\right)}{1+\dfrac{8\alpha t'}{w_l{}^2}} \\
&\times\left\{erf\left(\frac{z}{2\sqrt{\alpha t'}}-\mu_a\sqrt{\alpha t'}\right)-erf\left(\frac{z-d}{2\sqrt{\alpha t'}}-\mu_a\sqrt{\alpha t'}\right)\right\}dt'
\end{aligned} \quad (2)
$$

where $\mu_a$ [cm$^{-1}$] is the absorption coefficient, α [m$^2$/sec] is the diffusivity, ρ=1030[kg/cm$^3$] is the tissue density, d[m] is the tissue thickness, $u(t)$ is a step function, τ is the pulse duration, $w_l$ [m] is 1/e$^2$ radius of the beam and $E_0$ is the power density [W/cm$^2$].

| **Laser Properties:** | |
|---|---|
| Radiant Exposure | 1-100 mJ/cm$^2$ for λ=1.855μm<br>0.05-50 mJ/cm$^2$ for λ=1.937μm |
| Wavelength (λ) | 1.855μm ,1.937μm |
| Fibre Diameter | 200 μm |
| Beam Diameter (on sample) | 310 μm for λ=1.855μm<br>200 μm for λ=1.937μm * |
| Pulse Durations | λ=1.855μm: 35μsec, 100μsec, 300μsec, 600μsec, 1msec<br>λ=1.937μm: 5 μsec, 10 μsec, 30 μsec, 100 μsec, 300 μsec |
| Absorption | 10 mm$^{-1}$- for λ=1.937μm<br>1.5 mm$^{-1}$ for λ=1.855μm |

**Table 1 Simulation parameters for the auditory nerve stimulation experiments**
*** The fibre was coupled to the tissue**

A numerical model of an electrically stimulated auditory nerve fibre presented by Bruce *et al.*[23, 24] was used as a biophysical model for testing the INS stimulation. This model is an extension of Hill's deterministic threshold model for action potential generation[25]. To capture the stochastic nature of discharge probabilities of auditory nerve fibres a noise voltage, $V_{noise}$, which has a Gaussian amplitude distribution is added to the threshold value. This extended model reliably predicted data collected from electrically stimulated cats auditory nerve fibres[23].

# 3. Results

## *3.1 Photo-thermal excitation with extrinsic absorbers*

Short pulses (0.5 to 5ms) of intense, highly-localized light patterns were generated using a high-rate computer-generated holography system[20] with a green laser light source (λ=532nm, Altechna, Lithuania). The light patterns accurately targeted black micro-particles, and nearby cells were observed to be activated by the resulting micro-thermal transients (Fig. 1b). Calcium transients were induced for pulses as short as 0.5msec with no apparent damage to the cell. A threshold power level required for neuronal activation was measured for the different pulse durations (Fig. 1c, n=35 cells, $\tau$=0.5, 1, 2, 3, 4 & 5 msec)

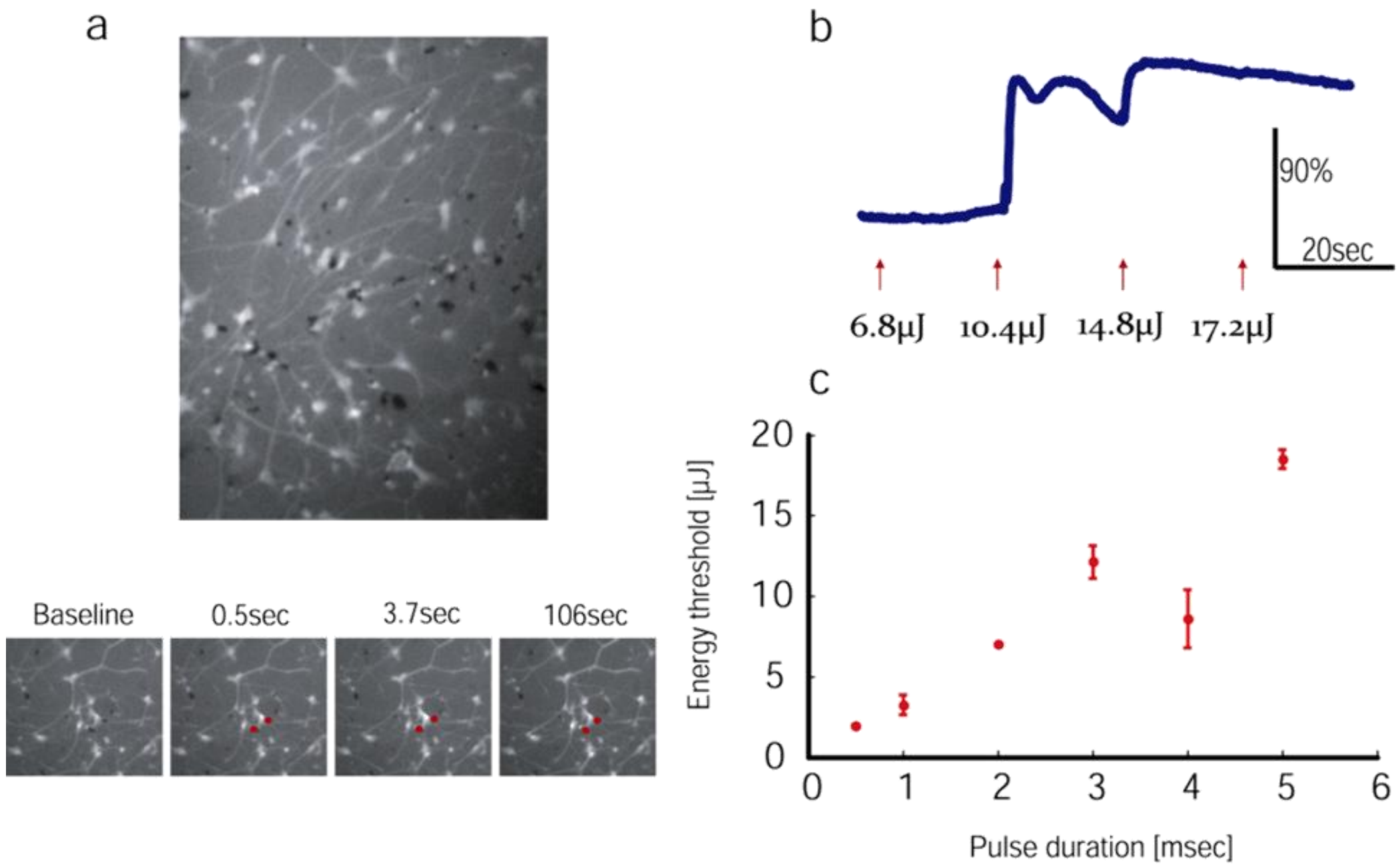

**Figure 1. *PAINTS* stimulation experiments. a. OGB-1 labelled cortical cell culture with photo-absorbers dispersed in the vicinity of the cells (top), a fluorescence change in cell adjacent to targeted photo-absorbers(red circles) following a 3msec pulse(bottom). b. The calcium transient trace induced by a 4msec pulse with increasing power for a cell adjacent to a targeted particle. c. Energy thresholds as a function of pulse width.**

### *3.2 PAINTS model*

To calculate the temperature transients elicited in the PAINTS experiments, the power thresholds measured in these experiments for the various pulse durations were substituted into Eq. 1. Figure 2 illustrates the resulting temperature transients (Fig. 2a) and their derivatives (Fig. 2b) for these parameters

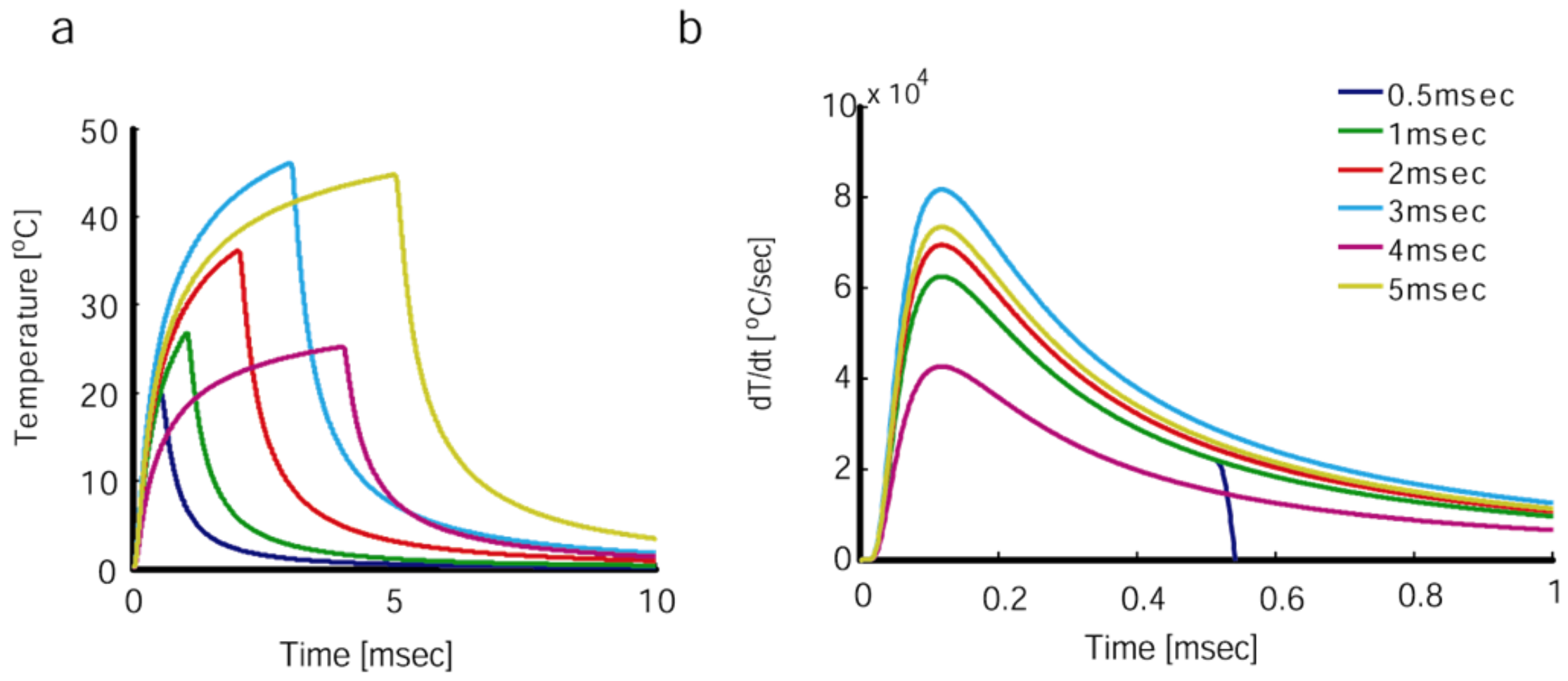

**Figure 2. Temperature transients temporal profile. a. Temperature transients calculated under the assumption of a discrete absorber, for various pulse duration and for power densities thresholds measured in the *PAINTS* experiments. b. The time derivative of the calculated temperature transients (negative temperature rates during cooling phase not shown, but note sharp transition to negative seen for 0.5 msec pulse).**

Simulation thresholds for the various pulse durations were estimated by inserting scaled temperature derivatives as somatic stimulation currents in the NEURON model. Action potentials were successfully elicited when the scaling factor crossed a certain threshold. Figure 3.a illustrates the resulting action potentials for increasing scale factors applied to a 0.5msec pulse - factors lower than the stimulation threshold fail to elicit an action potential. The predicted energy thresholds were calculated by multiplying the simulation power by the threshold scale factor and by the pulse duration, and were compared to the empirical data. Figure3.b demonstrates the good fit between empirical data and predicted energies under the temperature–rate model.

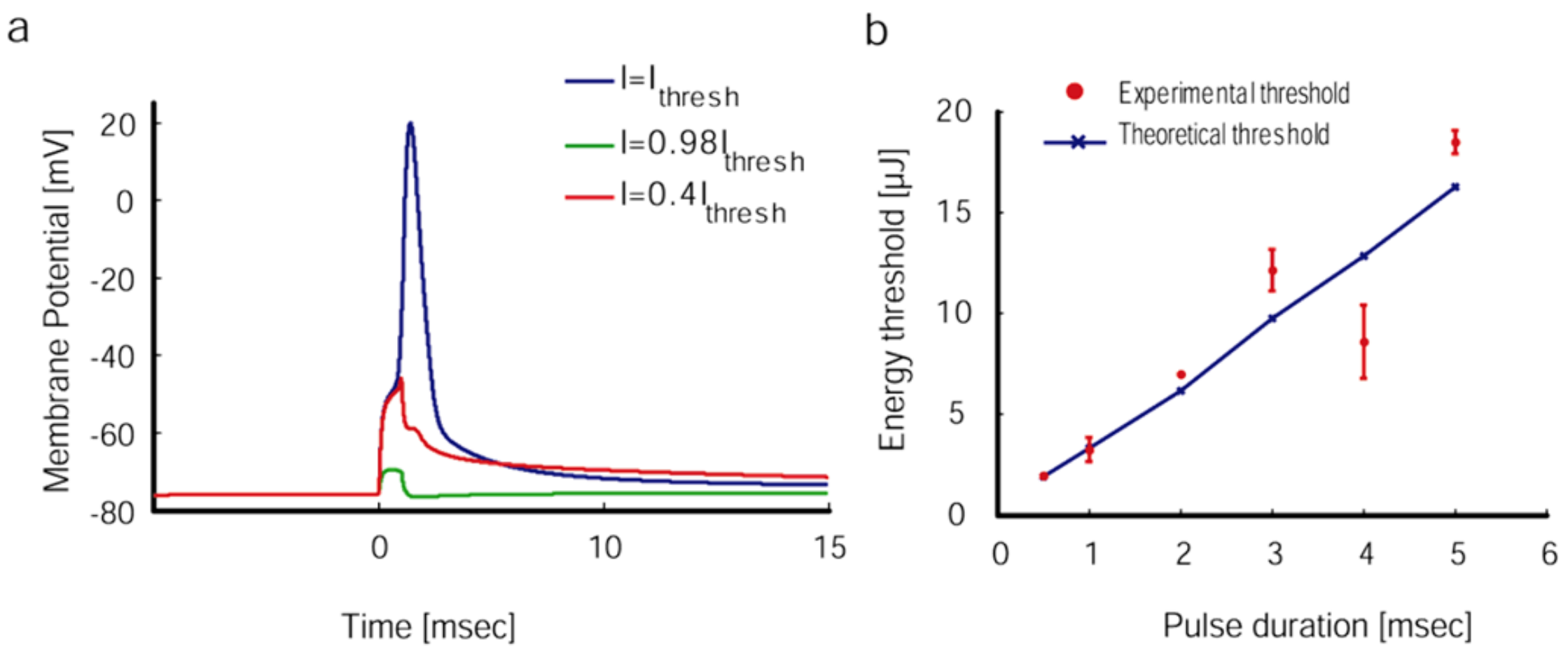

**Figure 3. Cortical cell culture computer modelling. a. membrane potential obtained from the NEURON model simulations for increasing scale factor for a 0.5 msec pulse. b. Energy thresholds predicted using the NEURON model for varying pulse durations (blue line) compared to the measured energy thresholds (red error bars)**

### *3.3 Auditory Nerve INS model*

The integration for calculating the temperature transient (Eq. 2) was solved for the parameters in Table 1. Figure 4 describes the spatial and temporal distributions of the temperature transient following a 300μsec pulse for both wavelengths used by Izzo *et al.*[18, 19]. The time derivative of the induced temperature transients for the various pulse durations used were calculated (Fig. 4.c). Note the different spatiotemporal behaviour of the temperature transients for the varying wavelength-dependent absorption.

Stimulation currents proportional to these temperature rates were injected into the biophysical model described above. A scaling constant was derived as the ratio $I_{thresh}$/ $I_{PT}$ between the threshold currents for auditory nerve stimulation $I_{thresh}$ (from the computer model), and the maximal value of the temperature derivative $I_{PT}$, calculated for a constant pulse energy of 1mJ/cm$^2$ for each pulse duration $\tau$. The threshold pulse energy is seen to be proportional to this ratio.

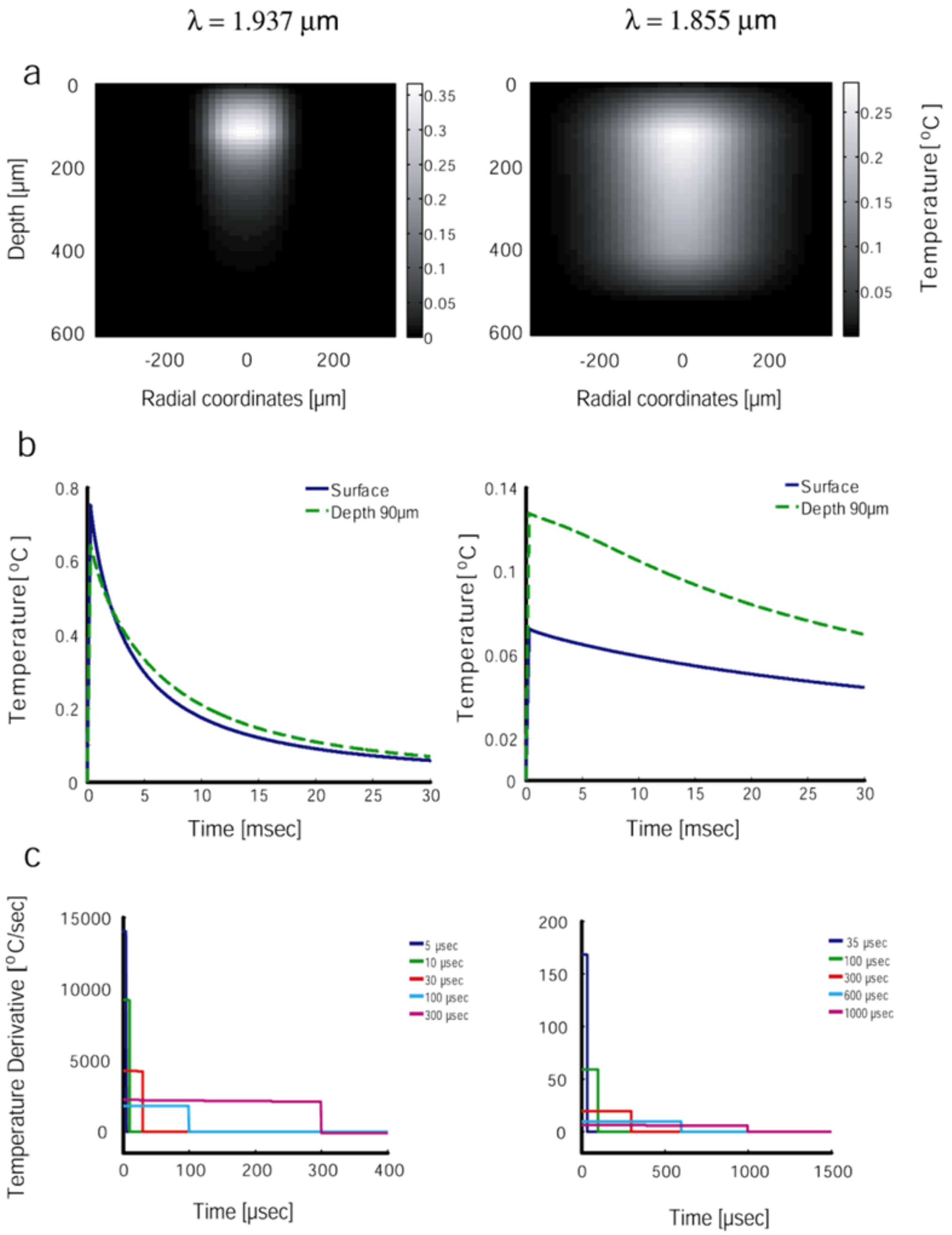

**Figure 4. Calculated temperature profiles for auditory INS experiments. a. Spatial distribution of temperature increase resulting from a 300μsec laser pulse, for a highly absorbed wavelength-1.937μm (left panel) and for a less absorbed wavelength 1.855μm (right panel) b. Temporal profile of temperature increase resulting from a 300μsec laser pulse, for a highly absorbed wavelength-1.937μm (left panel) and for a less absorbed wavelength -1.855μm (right panel). c. Temperature derivative for the various pulse durations used by Izzo for λ=1.937 μm (left panel) and λ=1.855 μm (right panel).**

Figure 5 compares between these theoretically predicted energy thresholds and the empirical data obtained by Izzo and colleagues.

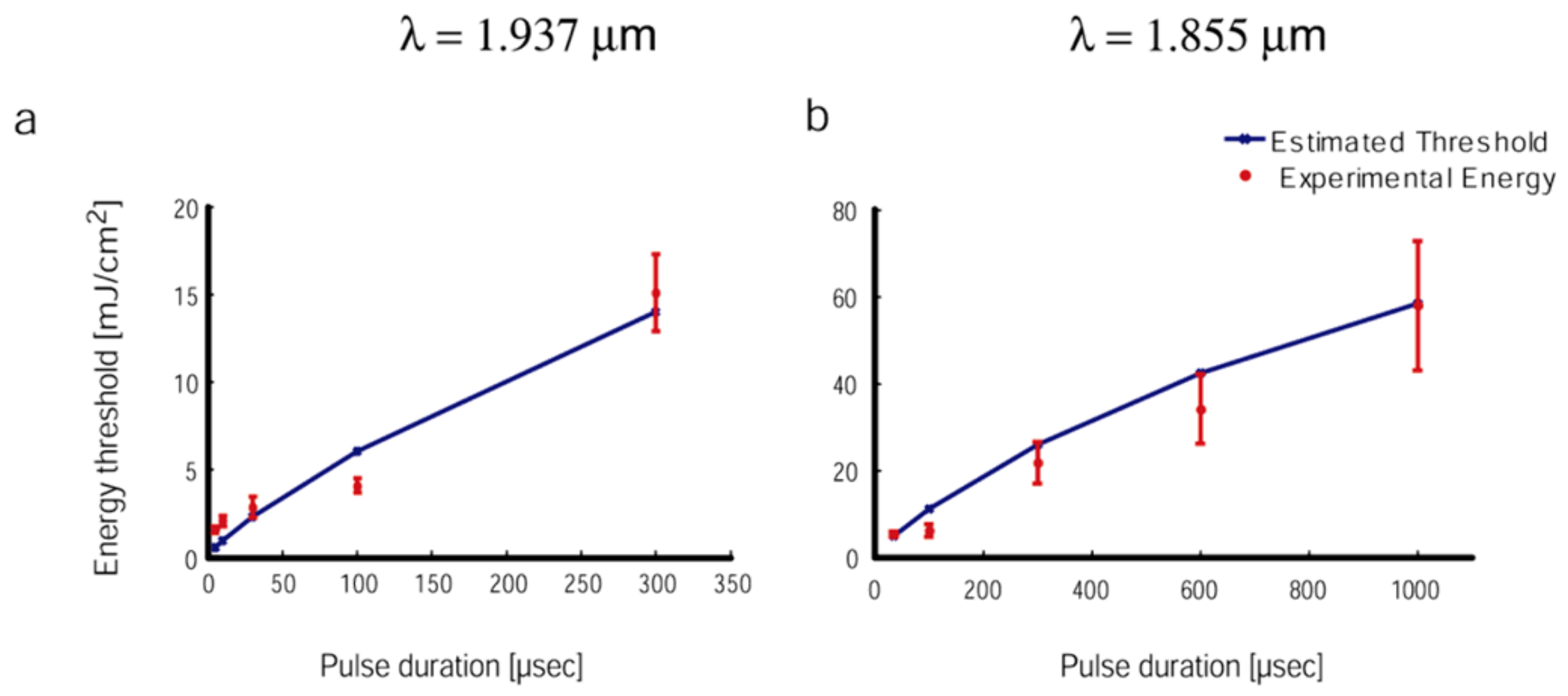

**Figure 5 Stimulation thresholds as a function of pulse duration. a. Estimated energy calculated using the biophysical and physical models compared to the stimulation thresholds obtained by Izzo et al for λ=1.937µm .b. Estimated energy calculated using the biophysical and physical models compared to the stimulation thresholds obtained by Izzo et al for λ=1.855µm**

# 4. Discussion

In this paper, we introduced and tested a temperature-rate based hypothesis for the quantitative description of the effects of photo-thermal stimulation on neural tissue. To test this hypothesis, we studied neuronal computer models where a thermally-induced trans-membrane stimulation current proportional to the rate of temperature change ($I \propto dT/dt$) was injected as a somatic stimulation current. The stimulation thresholds obtained in these simulations were found to be in very good agreement with empirical photo-thermal activation threshold results. This was shown both for *PAINTS* thresholds measured here in a cortical cell culture with black microparticle photo-absorbers (Fig. 3) as well as for data obtained by Izzo, Richter and colleagues[18, 19] in their experiments of auditory neuronal stimulations (Fig. 5). Another way of looking at this is by examining how constant is the scaling ratio $I_{thresh}/(dT/dt)_{max}$ for the different pulse durations reported in each experiment. The coefficient of variation (standard deviation/mean) of these ratios in the experimental data in figs. 3b and 5b is 0.25 and 0.28 respectively.

What are the possible underlying mechanism driving such temperature-rate effects?

First, it appears that these effects are probably not mediated through temperature effects on channel rate kinetics (Q10). To examine this issue, we incorporated the Q10 factor effect on the ionic channels transition rates and conductance values into our cortical cell model, and observed that this failed to initiate action potentials (data not shown).

Second, it appears that these effects are exactly opposite to the temperature-rate proportional hyperpolarization predicted by Barnes' analysis of the Potassium Nernst potential. Although this result inspired our search for a temperature-rate dependent effect, it is inconsistent with intracellular measurements during INS[13], and is not generally consistent with the observed excitations.

Third, it was argued[13] that the uniformity of photo-thermal excitation and the negative reversal potential observed appear to rule out TRP channels as an underlying mechanism, suggesting that the mechanism is a more general and generic effect on the cells' membrane. Arvanitaki and Chalazonitis[26] also highlighted the role of membrane impedance changes during their studies of photo-thermal excitation of aplysia neurons.

Finally, examining the membrane equation:

$$I_m = \frac{dV_m}{dt} C_m + V_m \frac{dC_m}{dt} + G_{Na}(V_m - V_{Na}) + G_K(V_m - V_{Na}) + G_{Cl}(V_m - V_{Cl})$$

it becomes apparent that a temperature-rate current can easily derive from temperature-dependent changes in the membrane's capacitance. Such changes could, e.g., result from rapid temperature-dependent changes in environmental pressure or in intramembrane hydrophobic forces[27], which could both potentially lead to rapid membrane narrowing and thus a depolarizing current (M. Plaksin and E. Kimmel, personal communications – interestingly, related effects are thought to be responsible for ultrasound neural stimulation[28]).

In conclusion, the empirical temperature-rate law presented here for photo-thermal neural stimulation could help direct future basic and applied studies of these phenomena. Under this model's assumptions, when a square INS pulse with duration much shorter than the thermal confinement time is applied, the injected current will approximate a square pulse of the same duration, proportional to pulse power. Thus, the injected *charge* will be proportional to *pulse energy*, and thresholds can potentially be derived from classical results for biophysical strength-duration relations[16].

# 5. Acknowledgements

The authors gratefully acknowledge the financial support of the European Research Council (starting grant #211055). N.F. was supported by a fellowship from the Israeli Ministry of Science and Technology. We would like to thank Michael Plaksin and Prof. Eitan Kimmel for productive discussions on physical effects on membranes.

# References

1. Yizhar, O., Fenno, L. E., Davidson, T. J., Mogri, M. & Deisseroth, K. Optogenetics in neural systems. Neuron 71, 9-34 (2011).
2. Wells, J. et al. Optical stimulation of neural tissue in vivo. Opt Lett 30, 504-6 (2005).
3. Richter, C. P., Matic, A. I., Wells, J. D., Jansen, E. D. & Walsh, J. T. Neural stimulation with optical radiation. Laser & Photonics Reviews 5, 68-80 (2010).
4. Teudt, I. U., Nevel, A. E., Izzo, A. D., Walsh, J. T., Jr. & Richter, C. P. Optical stimulation of the facial nerve: a new monitoring technique? Laryngoscope 117, 1641-7 (2007).
5. Izzo, A. D., Richter, C. P., Jansen, E. D. & Walsh, J. T., Jr. Laser stimulation of the auditory nerve. Lasers Surg Med 38, 745-53 (2006).
6. Harris, D. M., Bierer, M., Wells, J. & Phillips, J. Optical nerve stimulation for a vestibular prosthesis. Proc. SPIE, 7180, 71800R (2009).
7. Lee, D. J., Hancock , K. E., Mukerji, S. & Brown, M. C. Optical stimulation of the central auditory system. Abstr. Assoc. Res. Otolaryngol, 32,314 (2009).
8. Cayce, J. M. et al. Optical Stimulation of the Central Nervous System in vitro. Biomedical Optics, BTuE5 (2008).
9. Farah, N., Matar, S., Marom, A., Golan, L. & Shoham, S. Photo-absorber based neural stimulation for an optical retinal prosthesis. in Assoc. Res. Vis. Opth. (ARVO) abstr. no. 3470 (2010).
10. Farah, N., Matar, S., Golan, L., Marom, A., Brosh, I. & Shoham, S. Holographic photo-absorber induced neuro-thermal stimulation (PAINTS). in Soc. for Neuroscience (SFN) abstr. no 204.03 (2011)
11. Reutsky-Gefen, I. et al. Holographic Patterned Photo-Stimulation of Neuronal Activity for Vision Restoration. (submitted).
12. Wells, J. et al. Biophysical mechanisms of transient optical stimulation of peripheral nerve. Biophys J 93, 2567-80 (2007).
13. Katz, E. J., Ilev, I. K., Krauthamer, V., Kim do, H. & Weinreich, D. Excitation of primary afferent neurons by near-infrared light in vitro. Neuroreport 21, 662-6 (2010).
14. Teudt, I. U., Maier, H., Richter, C. P. & Kral, A. Acoustic events and "optophonic" cochlear responses induced by pulsed near-infrared laser. IEEE Trans Biomed Eng 58, 1648-55 (2011).
15. Sjulson, L. & Miesenbock, G. Photocontrol of neural activity: biophysical mechanisms and performance in vivo. Chem Rev 108, 1588-602 (2008).

16. Durand, D. M. in Biomedical Engineering handbook (ed. Bronzino, J. D.) (Boca Raton: CRC Press LLC,, 2000).
17. Barnes, F. S. Cell membrane temperature rate sensitivity predicted from the Nernst equation. Bioelectromagnetics 5, 113-5 (1984).
18. Izzo, A. D. et al. Optical parameter variability in laser nerve stimulation: a study of pulse duration, repetition rate, and wavelength. IEEE Trans Biomed Eng 54, 1108-14 (2007).
19. Izzo, A. D. et al. Laser stimulation of auditory neurons: effect of shorter pulse duration and penetration depth. Biophys J 94, 3159-66 (2008).
20. Golan, L., Reutsky, I., Farah, N. & Shoham, S. Design and characteristics of holographic neural photo-stimulation systems. J Neural Eng 6, 066004 (2009).
21. Roider, J. & &Birngruber, R. in Optical-Thermal Response of Laser-Irradiated Tissue (eds. Welch, A. J. & Martin, M. J. C.) (Plenum Press, New York, 1995).
22. Fleidervish, I. A., Lasser-Ross, N., Gutnick, M. J. & Ross, W. N. Na+ imaging reveals little difference in action potential-evoked Na+ influx between axon and soma. Nat Neurosci 13, 852-60 (2010).
23. Bruce, I. C. et al. A stochastic model of the electrically stimulated auditory nerve: single-pulse response. IEEE Trans Biomed Eng 46, 617-29 (1999).
24. Bruce, I. C., White, M. W., Irlicht, L. S., O'Leary, S. J. & Clark, G. M. The effects of stochastic neural activity in a model predicting intensity perception with cochlear implants: low-rate stimulation. IEEE Trans Biomed Eng 46, 1393-404 (1999).
25. Hill, A. V. Excitation and Accommodation in Nerve. Proceedings of the Royal Society of London. Series B - Biological Sciences 119, 305-355 (1936).
26. Chalazonitis, N., Romey, G. & Arvanitaki, A. [Resistance of the neuromembrane as a function of the temperature (neurons of Aplysia and of Helix)]. C R Seances Soc Biol Fil 161, 1625-8 (1967).
27. Widom, B., Bhimalapuram, P. & Koga, K. The hydrophobic effect. Physical Chemistry Chemical Physics 5, 3085-3093 (2003).
28. Krasovitski, B., Frenkel, V., Shoham, S. & Kimmel, E. Intramembrane cavitation as a unifying mechanism for ultrasound-induced bioeffects. Proceedings of the National Academy of Sciences (2011).